\documentclass[
 reprint,aps,showpacs,superscriptaddress,amsmath,amssymb,floatfix
]{revtex4-2}

\usepackage{graphicx}
\usepackage{mathrsfs}
\usepackage{tikz}
\usetikzlibrary{decorations.pathmorphing,arrows.meta,bending,positioning}
\usepackage{bm}

\newcommand{\scri}{\mathscr{I}}
\newcommand{\tg}{\tilde{g}}
\newcommand{\tC}{\widetilde{C}}
\newcommand{\tM}{\widetilde{M}}
\newcommand{\tP}{\widetilde{P}}

\newcommand*{\hnabla}{\widehat\nabla}

\newcommand*{\hP}{\widehat P}
\newcommand*{\hGammae}[3]{{\widehat\Gamma_{\bm{#1}\bm{#2}}{}^{\bm{#3}}}}

\begin{document}

\title{Fully non-linear gravitational wave simulations from past to future null-infinity}

\author{Jörg Frauendiener}
\email{joerg.frauendiener@otago.ac.nz}
\affiliation{Department of Mathematics and Statistics, University of Otago, Dunedin 9016, New Zealand}
\author{Chris Stevens}
\email{chris.stevens@canterbury.ac.nz}
\affiliation{School of Mathematics and Statistics,
  University of Canterbury, Christchurch 8041, New Zealand}
\author{Sebenele Thwala}
\email{sebenele.thwala@pg.canterbury.ac.nz}
\affiliation{School of Mathematics and Statistics,
  University of Canterbury, Christchurch 8041, New Zealand}

\date{\today}

\begin{abstract}
We present the first numerical simulations of asymptotically flat space-times whose computational domain includes past and future null-infinity. As an application, we explore the scattering of a gravitational wave in a black hole space-time. The study is conducted in the fully non-linear regime using an initial boundary value problem formulation of Friedrich's generalised conformal field equations. We calculate the Bondi energy, Bondi news and gravitational information at both past and future null-infinity and discuss their relationships.
\end{abstract}


\maketitle


\section{Introduction}

The first rigorous studies of gravitational asymptotics of isolated systems were conducted in the 1960s in asymptotically flat space-times, beginning in axisymmetry \cite{bondi1962gravitational} and soon extended to the general case \cite{sachs1962gravitational}. A global concept of mass associated with future-directed light-cones was introduced, now known as the \emph{Bondi mass}, allowing for the proof that gravitational waves carry energy and are a physical consequence of Einstein's theory of general relativity. Shortly thereafter, Penrose developed a conformal compactification technique for isolated systems, which introduced a conformal boundary, $\scri$, representing the points at infinity of the physical space-time in an elegant way \cite{penrose1963asymptotic,penrose1964light,penrose1965zero}. With the help of this boundary, one can unambiguously define asymptotic quantities such as gravitational radiation, energy, mass, and momentum without performing asymptotic expansions~\cite{penrose1984spinors}. For asymptotically flat space-times, the null hypersurface known as \emph{future null-infinity} $\scri^+$ consists of all the (idealised) endpoints of future-directed null geodesics. It plays a fundamental role in various aspects of gravitational physics. For example, null-infinity provides the backdrop against which outgoing gravitational waveforms emitted by binary systems are described in a well-defined way and in relation to which they can be approximated by numerical relativity codes, usually without having direct access to the surface itself \cite{bishop2016extraction}. These theoretical waveforms are then used together with LIGO observations to estimate parameters of the binary system \cite{abbott2016properties}.

Questions about the global structure of space-times have gained renewed theoretical interest recently as well due to an apparent close relationship between the asymptotic structure of space-time, symmetries, and conservation laws with the infrared structure and soft theorems \cite{strominger2018lectures} of quantum field theories.

Scattering is a well-known and instrumental concept in classical and quantum physics. Going back to the first experiment by E.~Rutherford, who discovered the structure of atoms by directing $\alpha$-particles onto gold foils, it was developed much further, both in experimental and theoretical directions. The ``scattering matrix'' has been proven to be a valuable tool in many aspects of quantum field theory. It describes the behaviour of a `system' in terms of its interaction with `particles'. It relates `ingoing states' approaching the system from the past with `outgoing states' receding from the system towards the future. These states are considered `asymptotically free', i.e., without interaction with the system in the limits of infinite distance in the past and in the future. These states are conventionally thought of as incoming and outgoing particles or waves. 

Considering electromagnetic or gravitational waves, this rough idea of scattering leads us directly to the concept of an asymptotically flat space-time with the asymptotic states corresponding to waves coming in from infinity, i.e., from~$\scri^-$ and dispersing out to infinity, i.e., to $\scri^+$. Thus, we are led to consider a problem where data are given at past null-infinity and where the evolution determines information at future null-infinity in the form of outgoing radiation. Since, in our context, null-infinity is a null hypersurface, this amounts to a characteristic initial value problem.

There is an active research area in mathematical relativity working toward the analytical treatment of the solution of this problem~\cite{nicolas2016conformal,mason2004conformal,joudioux2012conformal}, but so far, the focus has been on simple cases. However, the numerical treatment of the problem is limited by the fact that the methods to solve standard Einstein equations, such as the BSSN formalism \cite{Baumgarte1998,Shibata1995,Nakamura1987}, do not allow direct access to the asymptotic regions, which are ``at infinity'' and therefore are not accessible with finite resources. Thus, if the scattering problem is to be treated properly, one needs to employ methods that go beyond the standard numerical approaches and incorporate null-infinity in one way or another. 


One possibility is the use of hyperboloidal hypersurfaces, i.e., space-like hypersurfaces extending out toward null-infinity and suitably spatially compactified so that $\scri$ is included as a boundary of the spatial extent of the computational domain. This approach is currently being tested~\cite{zenginoglu2008hyperboloidal,gasperin2020hyperboloidal}, but it is difficult to incorporate both past and future null-infinity simultaneously. 

The same statement applies to characteristic codes, i.e., formulations of the Einstein equations based on a foliation of space-time by null hypersurfaces such as described in~\cite{bishop1993numerical,bishop1996cauchy,bishop1999cauchy}. Again, these approaches allow access to either future or past null-infinity but not both.

At the moment, the most promising methods for treating the scattering problem are those which rescale the physical problem such that both null infinities can be included in the computational domain. In this letter, we present the first attempt at the numerical treatment of the scattering problem, which is based on the Generalized Conformal Field Equations (GCFE) developed by H. Friedrich~\cite{friedrich1998gravitational} and implemented numerically by the authors~\cite{beyer2016numerical,Frauendiener_2021, frauendiener2023non, frauendiener2024non}.

Having access to both null infinities simultaneously allows us to relate the asymptotic quantities, such as the Bondi-Sachs energy-momentum, angular momentum or even the NP constants \cite{newman1968new}, defined on either null-infinity and, in this way, obtain statements about the global structure of space-time, and the interaction of the incoming waves which can be either gravitational or electromagnetic or both with `the system'. By this term, we mean, in our context, the space-time which would develop without any incoming radiation. This could be a flat Minkowski space or some isolated system, such as a static or stationary black hole from the Kerr-Newman family, or --- at least in principle --- even more complicated cases.

The primary purpose of this letter is to present, for the first time, a numerical formalism that includes both $\scri^+$ and $\scri^-$ in the computational domain and use it toward the treatment of the scattering problem. The letter is structured as follows: In Sec.~\ref{sec:setup}, we present an overview of the GCFE and their numerical implementation as an initial boundary value problem (IBVP). In Sec.~\ref{sec:results}, the Bondi energy, Bondi news and gravitational data on both $\scri^\pm$ are computed for the case of a gravitational wave impinging on a Schwarzschild black hole in the fully non-linear regime in axisymmetry. Various relationships between these data sets are explored. In Sec.~\ref{sec:discussion}, the results are discussed, and future work is laid out.

Our conventions follow those of \cite{penrose1984spinors}. In particular, we adopt units with $c=G=1$. Throughout this work, we assume that the cosmological constant vanishes.

\section{Equations and numerical setup} 
\label{sec:setup}
\subsection{The equations}
Here, we summarise the main properties of the GCFE and their numerical implementation, referring the reader to Appendix~\ref{app:eqs} and \cite{beyer2017a} for details. 

Roughly speaking, the GCFE are a regular reformulation of the (vacuum) Einstein equations on \emph{physical} space-time $(\tM,\tg)$ for the \emph{physical} metric $\tg$ in terms of an unphysical metric $g$ on a conformal space-time $(M,g)$. The two metrics are conformally equivalent, i.e., there exists a conformal factor $\Theta$ on $M$ such that $g = \Theta^2 \tg$ and such that $\Theta>0$ on $\tM$, which is thought of as embedded in $M$. The set $\scri$ on which $\Theta$ vanishes with a non-vanishing gradient is a null hypersurface in $M$ interpreted as the conformal boundary of $\tM$ in $M$.

Instead of directly addressing the physical metric $\tg$, the GCFE formulates the requirement of $\tg$ being a vacuum metric in terms of a pair $(\Theta, g)$. The two formulations are equivalent in the sense that a solution of one gives a solution of the other. 

The main ingredient in our simulation is the conformal Gauß gauge, which is discussed in more detail below. It has tremendous simplifications but also some disadvantages that we will elaborate on in the conclusion. The main advantage of using this gauge is the fact that the subsystem for the gravitational field is a symmetric hyperbolic system. At the same time, all other equations are reduced to transport equations.

\subsection{Initial data, boundary data}

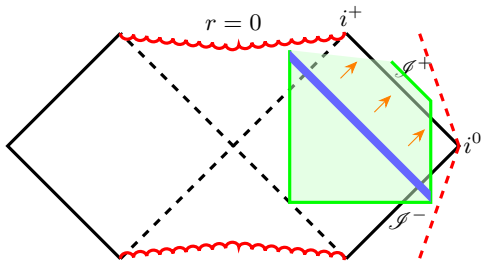
\begin{figure}
    \begin{center}
      \begin{tikzpicture}[very thick, decoration = {bent,amplitude=-5}, scale=0.75]
        \draw[dashed] (2,-2) -- (0,0)--(2,2);
        \draw (2,2)--(4,0) -- (2,-2);
        \draw[dashed] (-2,-2) -- (0,0)--(-2,2);
        \draw (-2,-2)--(-4,0) -- (-2,2);
        \draw[red,decoration={bumps,amplitude=-2},decorate] (-2,2) -- (0, 1.8) -- (2,2);
        \draw[red,decoration={bumps,amplitude=2},decorate] (-2,-2) -- (0, -1.8)-- (2,-2);

        \node at (4.25, 0.05) {$i^0$}; 

        \node at (2.1, 2.3) {$i^+$};

        \node at (3.2, 1.4) {$\mathscr{I}^+$};

        \node at (3.1, -1.3) {$\mathscr{I}^-$};

        \node at (0, 2.2) {$r=0$};

        \draw[draw  = green, very thick, fill=green!20!white, fill opacity=0.5] 
                        (1,1.7) -- (1,-1) -- (3.5,-1) -- (3.5,0.8) -- (2.8,1.5);

        \fill[ fill=blue!60, fill opacity=1.0] 
                        (1,1.5) -- (3.5,-1) -- (3.5,-0.8) -- (1,1.7) -- (1,1.5);

        \draw[dashed, red] (4,0) -- (3.3,-2);
        \draw[dashed, red] (4,0) -- (3.3,2);

        \draw[thin,-{Stealth[length=5pt,width=4pt]}, draw=orange]  (3.1,0) -- (3.4,0.3);
        \draw[thin,-{Stealth[length=5pt,width=4pt]}, draw=orange]  (2.5,0.6) -- (2.8,0.9);
        \draw[thin,-{Stealth[length=5pt,width=4pt]}, draw=orange]  (1.9,1.2) -- (2.2,1.5);
      \end{tikzpicture}
    \end{center}
    \caption{The computational domain (green shaded region) drawn over the conformal diagram of the Schwarzschild space-time. The blue-shaded strip corresponds to the path of an ingoing gravitational wave of compact support on the outer boundary. The orange arrows represent the outgoing gravitational radiation caused by non-linear back-reaction. The red dashed curves through $i^0$ represent the location where the hyperbolicity of the spin-2 subsystem is lost.}\label{fig:computationaldomain}
  \end{figure}

We aim to describe the scattering of gravitational waves at an initially static, i.e., Schwarzschild, black hole. Thus, we give initial data corresponding to exact Schwarzschild data on a hyperboloidal hypersurface $\Sigma$, which intersects $\scri^-$. The data essentially consist of the induced metric and the extrinsic curvature on $\Sigma$ from which initial values for all other quantities can be calculated. Given the hypersurface, one could compute these data exactly from an appropriate form of the metric for the Schwarzschild space-time. However, here we opt for a different approach in order to make use of the time-symmetry of the problem. We specify initial data induced from the Schwarzschild metric in isotropic coordinates on a hypersurface $\Sigma_0$ of time symmetry, i.e., with vanishing extrinsic curvature, and numerically evolve them backwards in time with the spherically symmetric GCFE, thus producing a hyperboloidal hypersurface $\Sigma$ with Schwarzschild data in the conformal Gauß gauge. Therefore, the conformal Gauß gauge is defined on $\Sigma_0$, which is given in these coordinates as $t=0$, while the hyperboloidal surface $\Sigma$ is given by $t=t_0<0$. In the absence of incoming gravitational waves, these data, when evolved into the future, will recreate the data on the previous hypersurface of time-symmetry. Furthermore, the time-slice which first intersects $\scri^+$ will be time-symmetric to $\Sigma$, i.e., at $t=-t_0$. Thus, using these initial data results in a computational domain as shown schematically in Fig.~\ref{fig:computationaldomain}. 

The key observation is that as the initial surface extends beyond $\scri^-$, we can prescribe boundary conditions that generate an ingoing gravitational wave propagating through $\scri^-$, implicitly setting the asymptotic characteristic initial data there. One needs to be careful, however, as the evolution equations lose hyperbolicity \footnote{See \cite{Beyer:2012} for further details.} on a particular time-like hypersurface outside $\scri$, represented as red-dashed curves shown in Fig.~\ref{fig:computationaldomain}. We find that there is sufficient room between $\scri^-$ and this surface to comfortably fit the wave profile before the outer boundary intersects $\scri^-$. The no-radiation boundary condition is then set for the remainder of the simulation on the outer boundary, and the GCFE are evolved up to and beyond $\scri^+$. This boundary condition approximates the behaviour near space-like infinity and does so more accurately the further out it is placed. The inner boundary position is chosen so that the gravitational radiation generated from the outer boundary reaches it \emph{inside} the apparent horizon; see Fig.~\ref{fig:computationaldomain}. In this case, the inner boundary is an outflow boundary; the only consideration required is numerical stability.

This setup allows us to propagate data from $\scri^-$ to $\scri^+$ in the fully non-linear regime, describing a finite burst of gravitational radiation impinging on a black hole. It has been found numerically that this choice of gauge continues to yield a regular frame up to and beyond $\scri^+$ even with the non-linear perturbations. The coordinates of the conformal Gauß gauge were set initially as the spatial isotropic Schwarzschild coordinates $(r,\theta,\phi)$, where $\phi$ can be suppressed due to axisymmetry. We choose the mass of the unperturbed black hole as  $m=0.5$. The initial surface $\Sigma$ is located at $t_0=-1.2$ and $r\in[0.25,1.25]$ there. The free boundary data for the ingoing characteristic mode $q_0$, corresponding to $\psi_0$ in a boundary adapted frame, on the outer boundary $r=r_o$ is chosen as
\begin{gather}
q_{0}(t,\theta)= \begin{cases}w(t,\theta) & -1.2 \leq t \leq -1.075\\ 0 & t > -1.075\end{cases}, \nonumber \\
w(t,\theta) = a\mathrm{i} \sqrt{\frac{2 \pi}{15}}\sin ^{8}(8 \pi (t+1.2))\,{ }_{2} Y_{20}(\theta), \label{eq:BCs}
\end{gather}
where the constant $a$ is the amplitude and ${}_2Y_{20}$ is the only axisymmetric (i.e., $\phi$ independent) $l=2$ spin-weighted spherical harmonic. This generates a ``bump'' for $q_0$ from the boundary with support in $t\in[-1.2,-1.075]$ that encapsulates the ingoing gravitational radiation. The purely imaginary wave profile corresponds to an axial mode injecting angular momentum into the system.

At the inner boundary, we impose the evolution equations using a one-sided stencil for the spatial derivative. The above choices completely specify the problem, and we now move to the numerical implementation.

\subsection{The numerical implementation}
We use the Python package COFFEE (COnFormal Field Equation Evolver) \cite{doulis2019a} that contains all the necessary numerical methods to evolve the system. Time integration is done using the standard fourth-order Runge-Kutta method, with radial derivatives calculated using the fourth-order summation-by-parts finite difference operator of Strand \cite{strand1994summation}. Angular derivatives are computed using a pseudo-spectral method, allowing for fast implementation of the $\eth$-derivatives \cite{beyer2016numerical}. Boundary conditions are imposed using the Simultaneous Approximation Terms (SAT) \cite{carpenter1994time} method, ensuring stability at the boundary. Computational speed is optimised through MPI and OpenMP parallelisation, and the computationally intensive parts are implemented in C. The code has operated successfully in full $3 + 1$ dimensions \cite{frauendiener2024non}; however, we assume axisymmetry to decrease computation time for our current work.

The radial and angular spatial dimensions are discretised into equidistant points. The adaptive time-step $\Delta t$ is fixed by $\Delta t = C\,\Delta r / K$, where $C$ is the Courant-Friedrichs-Lewy number taken to be $0.2$, $\Delta r$ is the radial step and $K$ is the maximum characteristic speed of the system on a given time-slice. We evolve the simulation up to $t=1.71$, and regridding is used to make the outer boundary hug $\scri^+$ on the outside.

\section{\label{sec:results}Results}

We can now relate ingoing physical information such as the energy to outgoing information in a well-defined manner. Checks of the correctness of the system and numerical implementation were carried out (see Appendix~\ref{app:calibration}) by checking the satisfaction of the constraints as well as both the Bondi-Sachs mass-loss on $\scri^+$ and the mass-gain on $\scri^-$. The Bondi energy and news, discussed in more detail in Appendix~\ref{app:bondicpts}, evaluated on $\scri^{\pm}$, are natural and relevant quantities that will be used below in exploring the scattering problem.

\subsection{The scattering problem}
\label{sec:scattering}

\begin{figure}[htb]
        \includegraphics[width=0.45\textwidth]{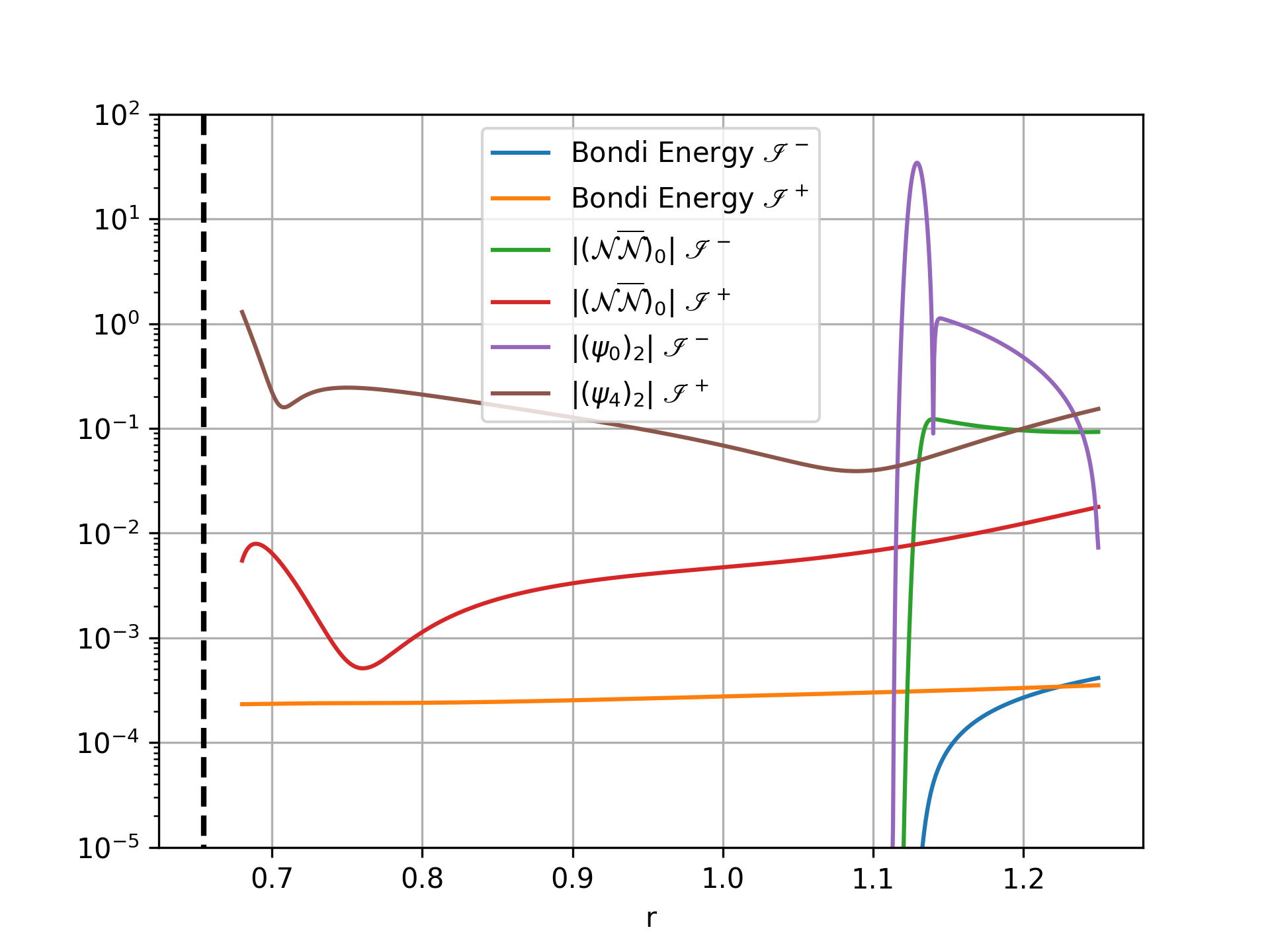}
    \caption{\label{fig:BE} The differences of relevant asymptotic quantities to their values in the unperturbed space-time along $\scri^-$ and $\scri^+$ for an initial wave amplitude $a=10$. The dashed line indicates future time-like infinity. Note that time runs towards the left (right) for quantities on $\scri^+$ ($\scri^-$) and that $(\psi_i)_l$ denotes the $(l,0)$-mode of $\psi_i$.}
\end{figure}

To explore the relationship between quantities on $\scri^-$  and $\scri^+$, we run simulations with ingoing wave amplitudes $a=1,2,5,10$ in Eq.~\eqref{eq:BCs}. Additional simulations were also performed with amplitude $a=10$, where the ingoing wave was chosen to have different frequencies. All simulations in this section were run with 3201 radial points, 65 angular points, and a CFL number of $C=0.2$. 

Fig.~\ref{fig:BE} presents the difference of the Bondi energy to the initial Schwarzschild mass $m=0.5$ and the energy flux, plotted against the radial coordinate for $a=10$. On $\scri^-$, the ingoing gravitational wave $\psi_0$ has a profile induced by the boundary data $q_0$ roughly in $r\in[1.11,1.14]$ -- the `bump'. Beyond this direct profile, there is a subsequent non-zero ``tail'' due to back-reaction during propagation from the outer boundary to $\scri^-$. The size of the tail is large immediately following the bump but then begins to decay quickly until the outer boundary intersects $\scri^-$ and enters the physical space-time. One can also see that the $l=0$ contribution to the flux $\mathcal{N}\bar{\mathcal{N}}$ is roughly constant after the initial bump, indicating that energy enters the physical space-time at an almost constant rate, even with the sharp decay of the ingoing wave. This shows a significant contribution from higher modes in $\psi_0$, which must have been generated by the back-reaction since the incoming boundary data was a pure $l=2$ mode. The constant non-zero flux is consistently reflected in the increasing Bondi energy. 

On $\scri^+$, we see a steady decline of the Bondi energy and the start of oscillating behaviour in $\psi_4$; a non-linear version of quasinormal ringing \cite{frauendiener2023non}. The oscillations in the Bondi energy are present but not visible in the plot. When the frequency of the ingoing wave was changed, the oscillation frequency of the Bondi energy along $\scri^+$ remained essentially the same. This shows that even in the non-linear regime, the oscillation frequency of the Bondi energy is primarily (and perhaps only) dependent on the angular mode of the perturbation.

We now turn to comparisons between fields on $\scri^-$ and $\scri^+$. It would be ideal, as in conformal scattering theory \cite{nicolas2024analytic}, to consider the full extent of $\scri^-$ and $\scri^+$. However, here, we can only look at finite portions of each. To keep things simple, we identify points on a cut of $\scri^-$ with points on a cut of $\scri^+$ if they lie on the same time-like conformal geodesic. This implies that points on identified cuts have the same value of $r$. This is possible because, in the conformal Gauß gauge, cuts are given by constant $t$ and $r$.

Our first scattering result concerns the ratio of outgoing to ingoing energy. We measure the ingoing (outgoing) energy $E_{in}$ ($E_{out}$) depending on the ingoing amplitude by integrating the ingoing (outgoing) energy flux. We find values for the ratio of the energies as given in Table~\ref{tab:Einout}.
\begin{table}[h]
    \centering
\begin{tabular}{c|cccc}
a & 1 & 2 & 5 & 10\\
\hline
\(E_{out}/E_{in}\) & 8.5\% & 8.8\% & 11.2\% & 20.2\%\\
\end{tabular}
    \caption{Ratio of outgoing to ingoing energy as a function of amplitude.}
    \label{tab:Einout}
\end{table}
This shows that the main part of the energy entering the system contributes to the Bondi energy. However, with increasing amplitude, the portion of energy converted into outgoing energy increases. This is again the effect of non-linear self-interaction of gravitational waves, and it also shows that space-time is a rather stiff fabric. It takes a significant amount of energy to excite "ripples".

\section{\label{sec:discussion}Discussion and future work}

We have performed the first fully non-linear numerical evolution of an asymptotically flat space-time from $\scri^-$ to $\scri^+$ using the case of a gravitational wave impinging onto a static black hole in axisymmetry. The Bondi energy was computed on both $\scri^-$ and $\scri^+$, the Bondi-Sachs mass-loss on $\scri^+$ and the analogous mass-gain on $\scri^-$ were verified, and the integral formulae were used as checks of correctness. The first quantitative relationships between asymptotic properties on $\scri^-$ and $\scri^+$ were analysed.

The setup defined here is very preliminary because we do not have a direct handle on the ingoing radiation on $\scri^-$. Instead, we indirectly specify this information in terms of data on a time-like boundary in the unphysical space-time.  Our future goal is to devise a method for the \emph{direct} imposition of asymptotic characteristic initial data on $\scri^-$. Furthermore, a fully global discussion of the scattering problem requires the inclusion of space-like infinity, which at this moment is not feasible due to unresolved issues of regularity~\cite{Friedrich:2018}. Both of these improvements require a substantial amount of analytical and practical development. However, the results presented in this letter mark a significant step forward in exploring the global properties of spacetime and establish a crucial link between numerical relativity and scattering problems in mathematical relativity.

\begin{acknowledgments}
        JF thanks the Isaac Newton Institute for Mathematical Sciences, Cambridge, for its support and hospitality during the program ``Twistor Theory,'' in which work on this paper was undertaken. CS thanks Chris Gordon, Jörg Hennig and Denis Pollney for feedback on early drafts of the letter. This work was supported by an EPSRC grant no EP/R014604/1 and by the Marsden Fund Council from government funding managed by the Royal Society Te Apārangi of New Zealand.
\end{acknowledgments}

\appendix

\section*{Appendices}
\subsection{The equations}\label{app:eqs}

Although the equations implemented in~\cite{beyer2017a} were obtained using the space-spinor formalism \cite{sommers1980space}, this aspect is not necessary for the following discussion, and we present the equations in the tensorial formalism for clarity.

The GCFE exploit the conformal freedom produced by introducing a Weyl connection $\hnabla{\nabla}_a$. The choice of this connection is parametrised by a smooth 1-form $b_a$, defined by the equation $\hnabla_a \tg_{bc}=-2b_a \tg_{bc}$. The GCFE, written in terms of an orthonormal frame $\{e_{\bm{a}}\}$ with coordinate components $c_{\bm{a}}^\mu = e_{\bm{a}}(x^\mu)$, are
\begin{subequations}\label{eq:GCFE}
  \begin{gather}
  	\label{eq:GCFE1}
    e_{\bm{a}}(c_{\bm{b}}^\mu) - e_{\bm{b}}(c_{\bm{a}}^\mu) = \hGammae{a}{b}{c} c_{\bm{c}}^\mu - \hGammae{b}{a}{c} c_{\bm{c}}^\mu, \\[8pt]
    \label{eq:GCFE2}
    e_{\bm{a}}(\hGammae{b}{c}{d}) - e_{\bm{b}}(\hGammae{a}{c}{d}) =
    \left(\hGammae{a}{b}{e} - \hGammae{b}{a}{e}
    \right) \hGammae{e}{c}{d}\nonumber \\
      - \hGammae{b}{c}{e} \hGammae{a}{e}{d} +
    \hGammae{a}{c}{e} \hGammae{b}{e}{d} \nonumber\\
    + \Theta K_{\bm{a}\bm{b}\bm{c}}{}^{\bm{d}} -
    2\eta_{\bm{c}[\bm{a}}\hP_{\bm{b}]}{}^{\bm{d}} +
    2\delta_{[\bm{a}}{}^{\bm{d}}\hP_{\bm{b}]\bm{c}} - 2
    \hP_{[\bm{a}\bm{b}]}\delta_{\bm{c}}{}^{\bm{d}},
    \\[8pt]
    \label{eq:GCFE3}
    \hnabla_a \hP_{bc} - \hnabla_b \hP_{ac} = b_e K_{abc}{}^e,\\[8pt]
    \label{eq:GCFE4}
    \hnabla_e K_{abc}{}^e = b_e K_{abc}{}^e,
  \end{gather}
\end{subequations}
where hatted quantities such as the Schouten tensor $\hP_{ab}$ are defined with respect to the Weyl connection and $K_{abc}{}^d = \Theta^{-1}C_{abc}{}^d = \Theta^{-1}\tC_{abc}{}^d$ is the so-called gravitational field tensor. The gauge freedom represented by the co-vector $b_a$ together with the other gauge freedoms (coordinates, frame, and conformal factor) can be canonically fixed in terms of the conformal structure by specifying a congruence of time-like conformal geodesics, governed by
\begin{align*}
    u^c \Tilde{\nabla}_cu^a & = -2b_cu^cu^a + u_cu^cb^a,\\
    u^c \Tilde{\nabla}_cb_a & = b_cu^cb_a  - \frac{1}{2}b_cb^cu_a - u^c\Tilde{P}_{ca},
\end{align*}
where $u^a$ represents the unit tangent vector to the time-like curves, $u^au_a = 1$, $\tP_{ab}$ is the Schouten tensor of the physical space-time, which vanishes due to the vacuum equations, and $\Tilde{\nabla}$ is the Levi-Civita connection of $\tg$.

By adapting an orthonormal frame and coordinates to these curves in analogy to the Gauß gauge, we obtain the \emph{conformal Gauß gauge}. This has the remarkable property that the conformal factor can be completely determined along the conformal geodesics in terms of initial data for $u^a$, $b_a$, $\Theta$, and the geometry of the initial time-slice.

The GCFE \eqref{eq:GCFE} written in the conformal Gauß gauge can be split into evolution and constraint equations. These systems have the following properties: first, the evolution equations propagate the constraints in the sense that they remain satisfied over time if they were satisfied initially \cite{friedrich1995einstein}, and, second, all quantities except for the gravitational field tensor $K_{abc}{}^d$ satisfy transport equations along the conformal geodesics, while $K_{abc}{}^d$ satisfies a first-order symmetric hyperbolic system. We refer to this as the spin-2 system. This property directly shows the wave character of the gravitational field and demonstrates the importance of the tensor $K_{abc}{}^d$. Its components are related to the Newman-Penrose scalars $\psi_0,\ldots,\psi_4$ defined with respect to an appropriate frame. They carry essential information about the gravitational radiation on $\scri^\pm$.

The GCFE have two propagating (complex) physical degrees of freedom, which can be identified with the complex scalars $\psi_0$ and $\psi_4$ in an adapted frame on the boundary of the computational domain. These must be fixed by appropriate boundary conditions. At first glance, the evolution system has two ingoing modes at each boundary. However, one of them is determined uniquely in terms of the other if one imposes the condition that the constraints are satisfied on the boundary; see \cite{beyer2017a} for details. We use a maximally dissipative boundary condition \cite{friedrich1999initial} to fix the remaining ingoing mode at the boundary.

\subsection{Calibration}\label{app:calibration}
To check the correctness of the equations and our numerical implementation, we confirmed that all constraint equations converged with resolution to the correct fourth-order both on slices of constant $t$ and intrinsic to $\scri^\pm$ in the situation studied here, in analogy to what has been done previously \cite{beyer2017a}. The maximum constraint violation on $\scri^+$ for the resolution used in Sec.~\ref{sec:scattering} is $10^{-6}$.

The Bondi-Sachs mass-loss on $\scri^+$ and what we call, in analogy, the \emph{mass-gain} on $\scri^-$ was also confirmed to be satisfied within a relative error of $\approx10^{-7}$ throughout the simulations. This was found by calculating the Bondi energy between two cuts, ($\mathcal{M}_1$) and a subsequent cut ($\mathcal{M}_2)$, and comparing the difference to the flux integral between the cuts by computing the residual of the mass-loss formula via
\begin{equation}\label{eq:massloss}
   \Delta = \frac{1}{\mathcal{M}_1}\left( \mathcal{M}_2 - \mathcal{M}_1 + \frac{1}{4\pi}\int_{\scri^2_1} U_0\mathcal{N}\bar{\mathcal{N}}\text{d}^3V\right),
\end{equation}
where the quantities in the integral are defined in the next section.

\subsection{Bondi components on $\scri^\pm$}\label{app:bondicpts}

As an example of relating global quantities on $\scri^\pm$, we look at the Bondi energy. We follow the formalism derived in \cite{frauendiener2021new}, which provides a conformally invariant expression for the components of the Bondi 4-momentum co-vector on $\scri^+$ in a general gauge, and here provide the analogous expression on $\scri^-$.

The Bondi energy on a cut of $\scri^-$ is defined as the integral over the cut of the product of the mass aspect $\mathcal{M}$ and the time-translation element $U_0$ of the BMS group in an adapted gauge
\begin{equation}
    \label{bondi}
    \mathcal{M}=\frac{1}{4 \pi} \int_{\mathcal{C}} U_0(\sigma' N+ \overline{\eth'}^{2}_c \sigma' - A \psi_{2}) A^{-1} \,\mathrm{d}S,
\end{equation}
where $\eth'_c$ is the conformally invariant GHP operator \cite{penrose1984spinors}, $A$ is a scale factor along the null generators and $N$ is related to the news $\mathcal{N}$ \cite{frauendiener2021new} and is given by
\begin{equation}
    N := \Phi_{02} - \rho\bar{\sigma}^{\prime}-\eth\bar{\tau}' + \bar{\tau}'^2,
\end{equation}
where $\Phi_{02}$ is a component related to the trace-free Ricci tensor \cite{penrose1984spinors}. These can then be written in terms of quantities known from the GCFE evolution. It is found in \cite{frauendiener2021new} that an appropriate time translation is given by $U_0 = \Omega$, where $\Omega$ is the conformal factor that rescales the particular cut of $\scri$ to the unit 2-sphere.

The news $\mathcal{N}$ for $\scri^+$ satisfies (on $\scri^+$) $\eth_c\mathcal{N} = A\psi_3$ and this equation is solved numerically for $\mathcal{N}$ and used in Eq.~\eqref{eq:massloss} as a non-trivial check of correctness of the Bondi energy.

\bibliography{PRL}
\end{document}